# Generation of single colour centers by focussed nitrogen implantation

J. Meijer[1] , B. Burchard[1], M. Domhan[2], C.Wittmann[2], T.Gaebel[2], I.Popa[2], F. Jelezko[2], and J. Wrachtrup[2]

[1] Central laboratory of ion beam and radionuclides, Ruhr Universität Bochum, Bochum Germany
[2] 3. Physikalisches Institut, Universität Stuttgart, Germany

**Abstract**

Single defect centers in diamond have been generated via nitrogen implantation. The defects have been investigated by single defect center fluorescence microscopy. Optical and EPR spectra unambiguously show that the produced defect is the nitrogen-vacancy colour center. An analysis of the nitrogen flux together with a determination of the number of nitrogen-vacancy centers yields that on average two 2 MeV nitrogen atoms need to be implanted per defect center.

Implantation of impurities for controlled doping of semiconductors is a routine technique nowadays. The increasing requirements on precision doping in nanoelectronics, however have lead to a number of technical affords in highly spatial resolved doping even of single impurities. Also novel techniques, like solid-state quantum information processing require the implantation of single dopant atoms into solids with high precision[5]. Recently, for example, substantial progress in the deposition of single phosphorous atoms into silicon has been made[15]. Here we report on the implantation of nitrogen ions into diamond to generate nitrogen-vacancy defect centers. Individual NV defect centers can be detected optically such that the successful implantation is verified easily[3].

The nitrogen vacancy defect in diamond consists out of a substitutional nitrogen atom with a next nearest neighbour vacancy. The defect is characterized by a strong optical absorption around 1.945 eV (637nm) and intense fluorescence[2]. The NV center has been well characterized by various spectroscopic techniques. The 637nm absorption is attributed to the negatively charged form of the defect center[12]. The six-electron model gives rise to a spin triplet ground state ($^3A$) and a first excited triplet state ($^3E$)[9]. Electron spin resonance experiments show a strong spin polarization upon optical illumination[8]. The fine structure parameters were measured to be $D$ = 2.88 GHz and $E$ = 0 GHz, showing that the NV center spin wavefunction is axially symmetric and mostly localized at the defect site itself[14]. Measurements of the hyperfine coupling support this picture. 70% of the spin density is localized at the three unpaired carbon atoms while the residual spin density is mostly restricted to the second coordination shell of carbon atoms. Previously the NV defect has been produced in nitrogen rich (type 1b) diamond by electron or neutron irradiation. In this procedure vacancies are created. Subsequent annealing at 600°C results in vacancy diffusion such that eventually stable NV centers (up to 2000 °C) are formed. This method has also been used to generate defect centers with high spatial accuracy by using a focussed beam of Ga ions to produce vacancies[10].

In contrast, the present approach generates NV centers by nitrogen implantation into very pure type *IIa* diamond (nitrogen content <0.1ppm)[4]. Fig. 1 shows a series of images where nitrogen vacancy defects have been generated by implanting nitrogen in such a way. A beam of 2 MeV $N^+$ ions with diameter 0.3μm has been imaged on the sample. The $N^+$ beam is produced by the dynamitron tandem accelerator in Bochum (DTL) and focused using a high-energy ion projector equipped with a 15 T superconducting solenoid lens [11]. The regular pattern of implantation sites visible in Fig. 1 has been generated by stepwise raster scanning the beam over the sample. For the low dose implantation the ion beam current is reduced down to a few ions/ sec controlled by a silicon barrier surface detector. Yet, the setup presently does not allow a *controlled* implantation of the single ions. For the low dose implantation, there is thus a certain spread of number of ions per spot, which is defined by Poisson distribution of ions.

STRIM calculations suggest that the ions should be implanted 1 μm below the surface. The lateral straggling is expected to be 0.5μm. Both values are in agreement with our experimental findings. For each implantation spot optical emission spectra have been taken to investigate the nature of the defects created. The optical emission spectra in Fig. 1b show that after nitrogen implantation and prior to annealing mostly neutral vacancies (GR1 emission) are formed. After annealing the GR1 emission is converted into a fluorescence emission spectrum characteristic for the NV colour center. In schemes where NV centers are created via production of vacancies, the calculation of the conversion efficiency, i.e. number of electrons versus NV centers produced relies on the knowledge of nitrogen present in the material. For most materials, especially with low nitrogen content this is a not accurately known quantity. In contrast, in the present approach the conversion efficiency, i.e. the number of implanted nitrogen versus NV centers created can be reliably estimated. For this the number of implanted nitrogen atoms is determined from the nitrogen flux and implantation time (see e.g. Fig. 1). The number of the NV centers produced can be determined either via the fluorescence intensity per spot or via analysis of the photon counting statistics. In Fig. 2a, similar to Fig. 1, a regular pattern of NV centers has been generated by nitrogen implantation. However, a much lower flux has been chosen such that at some implantation spots no NV center has been generated. We estimate that on average two nitrogen atoms are implanted per spot. To determine the number of NV centers per spot, the photon statistics of the spot emission is evaluated. For this, the normalized second-order autocorrelation function $g^{(2)}(\tau)=\frac{\langle I(t)\cdot I(t+\tau)\rangle}{\langle I(t)^2\rangle}$ was measured. The measured traces, see Fig. 2b have been corrected for background according to reference[1]. In cases where the emission stems from a single NV center $g^{(2)}(\tau)$ should drop to zero for $\tau$ = 0. This phenomenon is commonly called antibunching and basically describes the fact that a single quantum system can only emit one photon at a time. Since $g^{(2)}(\tau)$ measures the conditional probability to find a second photon after at time $\tau$ if the first one has been emitted at time zero, $g^{(2)}(\tau)$ is expected to drop to zero for very small $\tau$ for a single quantum system. In general, if $N$ defect centers are optically excited, the autocorrelation function is given by[7]:

$$g^{(2)}(\tau)=1+\frac{1}{N}.\left(C_1\cdot e^{-\left|\tau/\tau_1\right|}+C_2\cdot e^{-\left|\tau/\tau_2\right|}\right)$$

Here $C_1$, , $\tau_1$ and $\tau_2$ are parameters which depend on internal relaxation rates of defect and on excitation intensity, and $C_2 = -1- C_1$. For most of the spots shown in Fig. 3 $g^{(2)}(0)$ indeed reaches values close to zero. For other spots the contrast is only 0.5 indicating the emission of two independent emitters. In few cases even a lower contrast, like 0.3 is found which points towards three defects per spot. Most of the points in Fig. 2a show photons emission behaviour, which is characteristic for a *single* NV center. Hence we conclude from the measurements that for 2 MeV implantation of nitrogen in diamond and subsequent annealing at 850 °C on average two nitrogen ions need to be implanted to generate a single colour center. Final prove that the defect centers generated are NV colour centers comes from optically detected magnetic resonance (ODMR). In this technique the fluorescence of the defect is observed while microwaves are scanned in the appropriate frequency range to induce magnetic dipole transitions between ground state spin sublevels. Fig. 3 shows the respective spectrum. The spin Hamiltonian used to simulate the spectrum is $H = \hat{S}D\hat{S} + \hat{S}A\hat{I} + P\left[I_z^2 - \frac{I^2}{3}\right]$, where $A$ is the hyperfine and $D$ the fine structure tensor. $P$ is the quadrupole coupling constant. For the spectrum in Fig. 3, $D$ is found to be 2880 MHz exactly as expected for the NV center. In addition the triplet line structure is attributed to hyperfine coupling of the electron spin to the $^{14}N$ nuclear spin (I=1) with hyperfine parameters $A_{\perp}$ = 2.1 MHz and $A_{||}$ = 2.3 MHz. The quadrupolar coupling constant is found to be $P$ = -5.04 MHz. The values are in close agreement with literature values[13] and we thus conclude that the fluorescence emission stems form a single NV center.

In summary we have demonstrated the generation of *single* NV colour centers via nitrogen implantation into diamond, which has a very low “natural” abundance of nitrogen. For 2 MeV implantation, on average two nitrogen ions need to be implanted to generate a single NV center. In future an improved positioning accuracy might allow for a spatially controlled implantation of defect centers. Because of the long dephasing times[6] usually measured for the defect center, quite small couplings might be detectable. For example a dephasing time of 0.06 ms will allow to spectrally resolve a coupling for two defect centers which are separated by 5nm.

**Acknowledgement**

The authors acknowledge helpful discussions with S. Prawer. The work has been supported by the DFG graduate college “Magnetische Resonanz”, the focussed programme

"Quanteninformationsverarbeitung", the EU via the project "QIPDDF-ROSES" and the Landestiftung BW via the program "Atomoptik".

**Figure captions**

Fig. 1 (a) Three different areas of a diamond sample implanted with different amounts of nitrogen. The pattern was generated by imaging a 2 MeV nitrogen beam with a beam diameter of 300nm on the diamond surface. The number of nitrogen ions implanted per spot is marked in the upper right corner. (b),(c) Fluorescence emission of a single spot before and after annealing.

Fig. 2. (a) Low-dose implantation pattern of nitrogen. Parameters are identical to Fig. 1 but average implantation dose is two nitrogen atoms per spot. (b) Normalized fluorescence intensity autocorrelation function of different spots from Fig. 2. The curves have been corrected for background contributions.

Fig. 3. Optically detected electron spin resonance spectrum of a single NV center. The smooth curve is a simulation of the spectrum by using the spin Hamilton operator $H$ described in the text. Inset shows energy level scheme of the triplet ground state of NV defect.

(a)

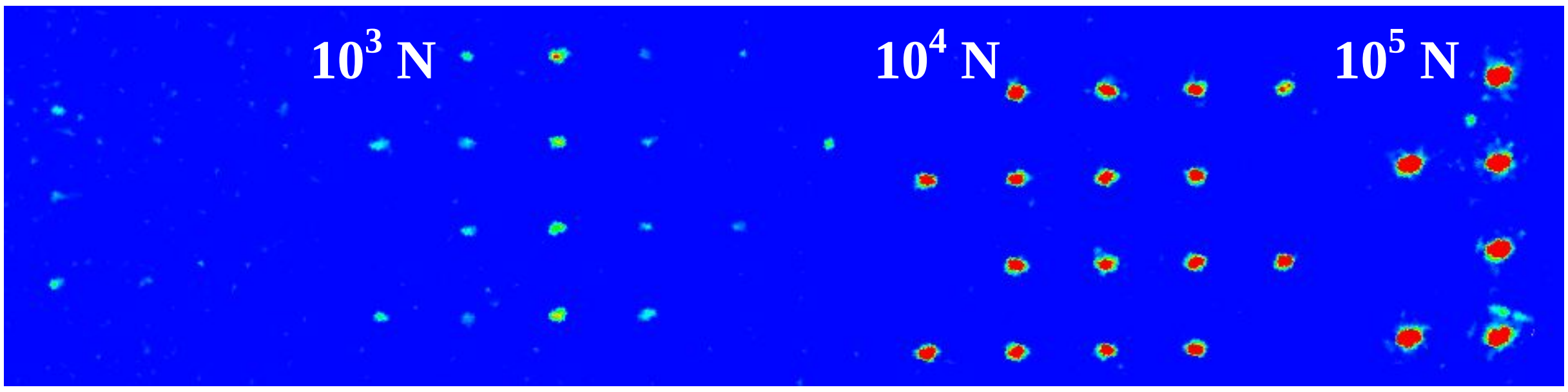
10^3 N
10^4 N
10^5 N

(b)

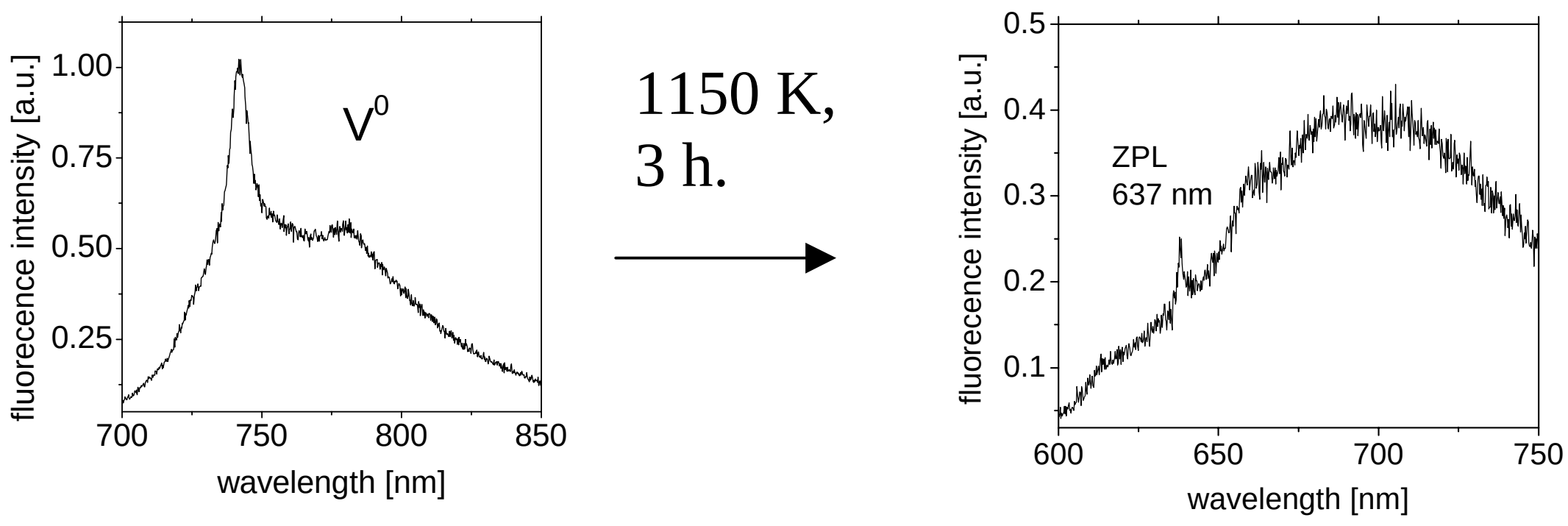

# Figure 1

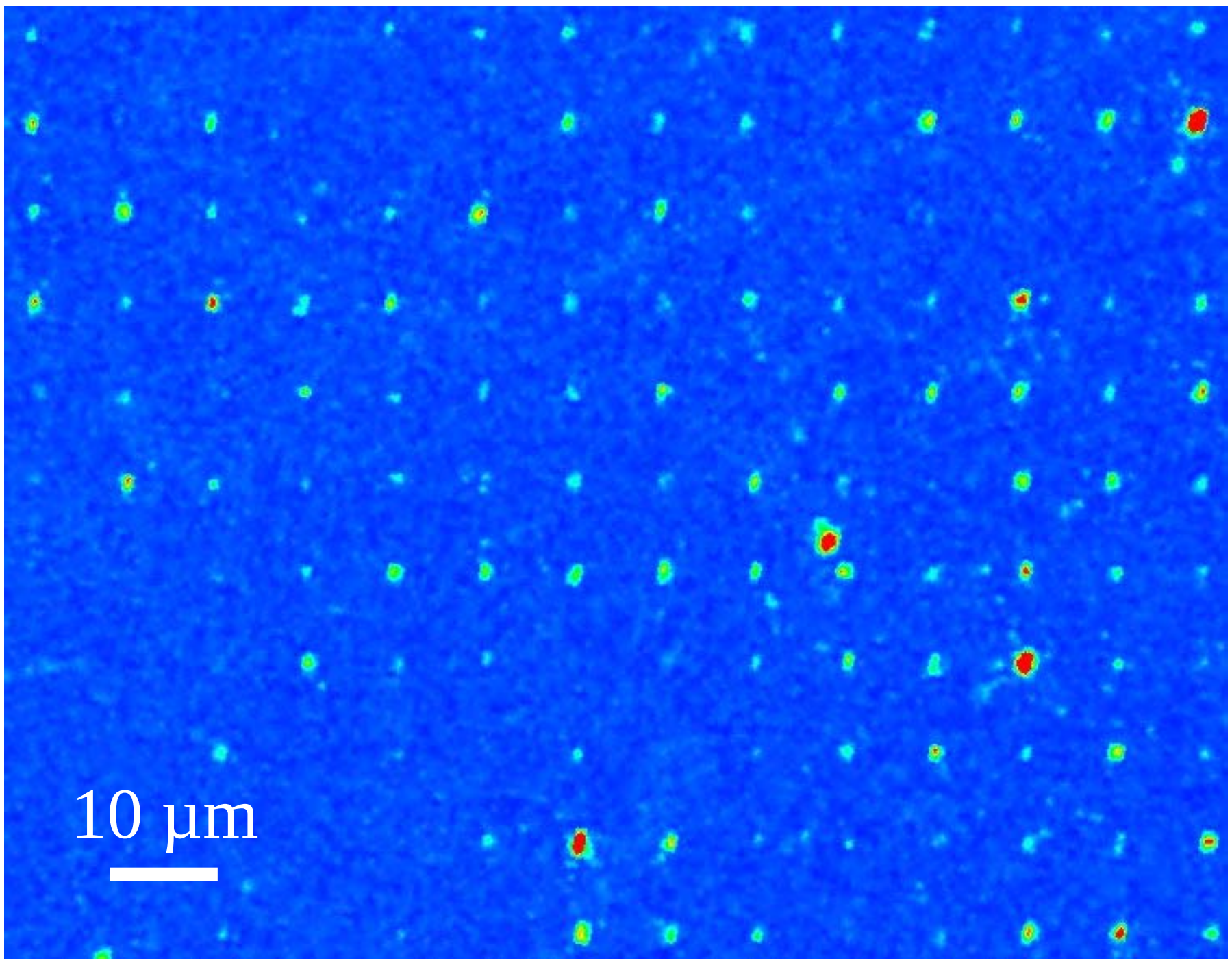

Figure 2 (a)

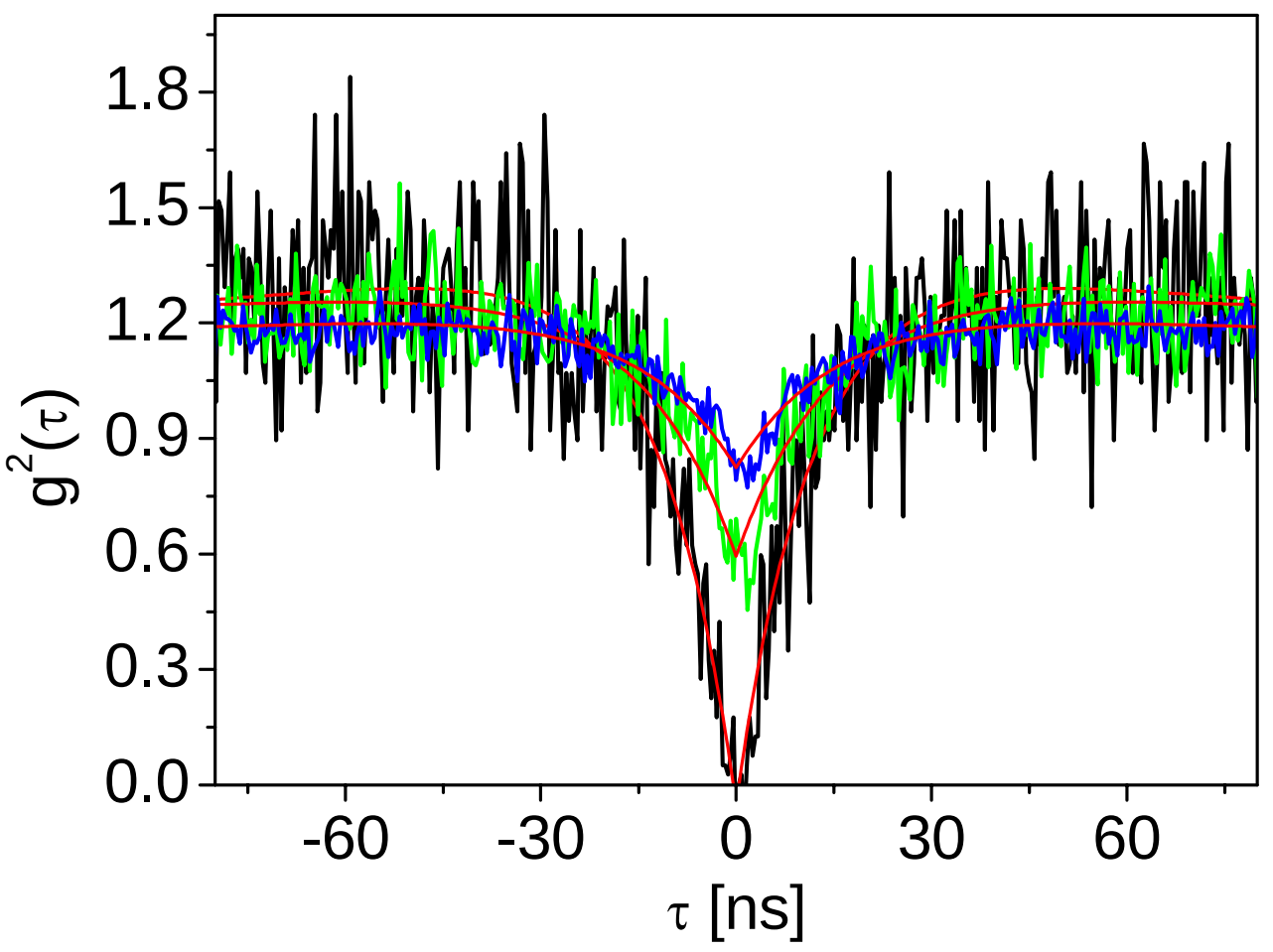

Figure 2(b)

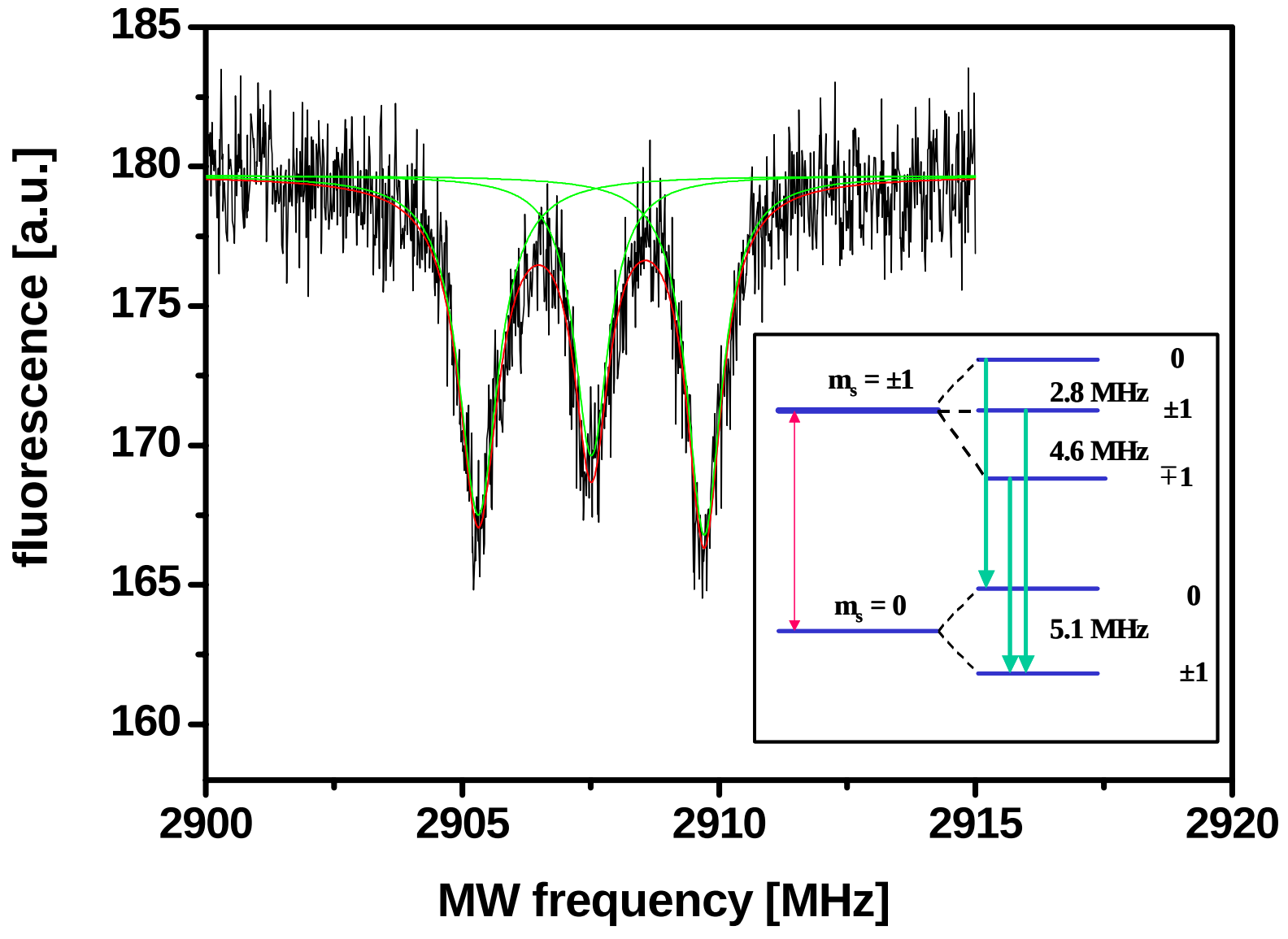

Figure 3

**Reference List**

1. Beveratos, A., Kuhn, S., Brouri, R., Gacoin, T., Poizat, J. P., and Grangier, P., European Physical Journal D **18**, 191 (2002).

2. Davies, G. and Hamer, M. F., Proceedings of the Royal Society of London A **384**, 285 (1976).

3. Gruber, A., Drabenstedt, A., Tietz, C., Fleury, L., Wrachtrup , J., and von Borczyskowski, C., Science **276**, 2012 (1997).

4. Kalish, R., UzanSaguy, C., Philosoph, B., Richter, V., Lagrange, J. P., Gheeraert, E., Deneuville, A., and Collins, A. T., Diamond and Related Materials **6**, 516 (1997).

5. Kane, B. E., Nature **393**, 133 (1998).

6. Kennedy, T. A., Colton, J. S., Butler, J. E., Linares, R. C., and Doering, P. J., Applied Physics Letters **83**, 4190 (2003).

7. Kurtsiefer, C., Mayer, S., Zarda, P., and Weinfurter, H., Physical Review Letters **85**, 290 (2000).

8. Loubser, J. H. N. and van Wyk, J. A., Reports on Progress in Physics **41**, 1201 (1978).

9. Martin J.P.D., Journal of Luminescence **81**, 237 (1999).

10. Martin, J., Wannemacher, R., Teichert, J., Bischoff, L., and Kohler, B., Applied Physics Letters **75**, 3096 (1999).

11. Meijer, J and Stephan, A, Nucl. Instr. Meth. B **188** , 9 (2002).

12. Mita, Y., Physical Review B **53** , 11360 (1996).

13. van Oort, E., Thesis/Dissertation, PhD Thesis, 1990.

14. van Oort, E., Stroomer, P., and Glasbeek, M., Physical Review B **42**, 8605 (1990).

15. Yang, C. Y., Jamieson, D. N., Pakes, C., Prawer, S., Dzurak, A., Stanley, F., Spizziri, P., Macks, L., Gauja, E., and Clark, R. G., Japanese Journal of Applied Physics Part 1-Regular Papers Short Notes & Review Papers **42**, 4124 (2003).